# Predicting EGFR Mutation in LUAD from Histopathological Whole-Slide Images Using Pretrained Foundation Model and Transfer Learning: An Indian Cohort Study

Sagar Singh Gwal[1*], Rajan[2*], Suyash Devgan[4], Shraddhanjali Satapathy[4], Abhishek Goyal[3], Nuruddin Mohammad Iqbal[4], Vivaan Jain[6], Prabhat Singh Mallik[5], Deepali Jain[4#], Ishaan Gupta[1#]

1. Department of Biochemical Engineering and Biotechnology, IIT Delhi, New Delhi, India
2. School of Interdisciplinary Research, IIT Delhi, New Delhi, India
3. School of Artificial Intelligence, IIT Delhi, New Delhi, India
4. Department of Pathology, AIIMS, New Delhi, India
5. Department of Medical Oncology, Dr. B.R.A.IRCH, AIIMS, New Delhi, India
6. Delhi Technological University, New Delhi, India

*\* authors contributed equally; # corresponding authors*

**Abstract**

Lung adenocarcinoma (LUAD) is a subtype of non-small cell lung cancer (NSCLC). LUAD with mutation in the EGFR gene accounts for approximately 46% of LUAD cases. Patients carrying EGFR mutations can be treated with specific tyrosine kinase inhibitors (TKIs). Hence, predicting EGFR mutation status can help in clinical decision making. H&E-stained whole slide imaging (WSI) is a routinely performed screening procedure for cancer staging and subtyping, especially affecting the Southeast Asian populations with significantly higher incidence of the mutation when compared to Caucasians (39-64% vs 7-22%). Recent progress in AI models has shown promising results in cancer detection and classification. In this study, we propose a deep learning (DL) framework built on vision transformers (ViT) based pathology foundation model and attention-based multiple instance learning (ABMIL) architecture to predict EGFR mutation status from H&E WSI. The developed pipeline was trained using data from an Indian cohort (170 WSI) and evaluated across two independent datasets: Internal test (30 WSI from Indian cohort) set, and an external test set from TCGA (86 WSI). The model shows consistent performance across both datasets, with AUCs of 0.933 (±0.010), and 0.965 (±0.015) for the internal and external test sets respectively. This proposed framework can be efficiently trained on small datasets, achieving superior performance as compared to several prior studies irrespective of training domain. The current study demonstrates the     feasibility of accurately predicting EGFR mutation status using routine pathology slides, particularly in resource-limited settings using foundation models and attention-based multiple instance learning.

## Introduction

Lung cancer is a significant public health concern in India, accounting for 7.8% of all cancer-related deaths (Noronha et al., 2024). Lung cancer can be broadly categorized into non-small cell lung cancer (NSCLC: 85%) and small cell lung cancer (SCLC: 15%). Lung adenocarcinoma (LUAD), the most common subtype of NSCLC, accounts for approximately 40% of all NSCLC cases. Approximately 68% of LUAD cases can be effectively treated with targeted therapy, which depends on identifying molecular subtypes. One such molecular subtype is an alteration in the epidermal growth factor receptor (EGFR) gene, which accounts for approximately 45.8% of LUAD cases (Jha et al., 2024). Clinical identification of LUAD molecular subtypes involves a combination of histopathological evaluation (e.g., H&E staining) and molecular diagnostics, such as polymerase chain reaction (PCR) and targeted next-generation sequencing (NGS). Though effective, these methods can be costly and time consuming, often taking days or even weeks to deliver results (Araki et al., 2023).

Advancements in artificial intelligence (AI) have revolutionized medical diagnosis, especially in image-based applications. AI models, such as machine learning (ML) and deep learning (DL), can identify intricate patterns often missed by human pathologists, thereby accelerating and automating the diagnostic procedure. AI enabled digital pathology can save significant time and reduce loss to follow-up of patients by enabling early detection through single tests, such as histopathology. In resource limited settings like India, precise molecular diagnostics technology like NGS can be expensive, and AI driven approaches can offer a cost effective, fast and robust alternative for patient triaging and diagnosis. Various AI models have been developed to identify and classify alterations in the EGFR gene using whole slide images (WSI) . These models are generally trained on large datasets, requiring substantial computational resources during training (Nguyen et al., 2025). Furthermore, limitations such as the lack of open-source models (including code and pretrained weights), limited access to large annotated datasets, low predictive power and lack of explainability have hindered testing, adoption and implementation of these models in new clinical setup serving a different new geography catering to patients with unique genetic profiles (D'Adderio & Bates, 2025; Magrabi et al., 2019). Pretrained foundation models can overcome these limitations by self-supervised feature extraction, enabling fast, robust, and generalizable feature representation learning from WSI. Vision Transformers (ViT) models like Prov-GigaPath are capable of capturing both fine-grained local features and long-range global dependencies through self-attention mechanisms leading to robust feature embedding (Xu et al., 2024). Additionally, a transfer-learning approach can significantly reduce training time for the final classification models (Kim et al., 2022). Transfer-learning is a powerful technique where a pretrained (trained on large, diverse datasets) model is repurposed for another related task by fine-tuning on relatively smaller datasets. This significantly reduces the data and computational requirements while improving the model's predictive power (Pan & Yang, 2010). These models are generally trained and evaluated on large-scale databases like

TCGA which has a highly skewed racial representation with an underrepresented Indian population (Spratt et al., 2016). Furthermore, cases of LUAD with EGFR mutations are more frequent in the Southeast Asian population (39-64%) compared to African (11-19%) and Caucasian (7-22%) population. India being a Southeast Asian country has substantially higher cases of LUAD with EGFR mutations (22-40%) as compared to their Western Counterparts (Chougule et al., 2013; Gutta et al., 2025; Malik et al., 2025). Also, the genetic makeup of the population, variability in imaging modality, sample preparation (e.g., staining, humidity and temperature fluctuation), and slide-scanning conditions can have significant impact on the appearance of WSI. Therefore, the adoption of domain generalization approaches is an imperative to validate AI models on region-specific datasets, such as those from India (Lafarge et al., 2017).

In a recent study by Campanella et al., 2025, the authors introduced a large and diverse dataset comprising histopathology images from multiple countries and institutions to address limitations related to domain and modality generalization. A ViT-based foundation model was employed for efficient feature embedding. The final model was trained and validated using 8461 whole-slide images, achieving an AUC of 0.847. Notably, the dataset was predominantly biased towards Caucasian origin (approx. 78%), with approx. 10% (850 WSI) accounting for the Asian descents. (Campanella et al., 2025). However, even this model, although available on open-source platforms, has issues with deployment. To overcome these challenges, we have developed a novel pipeline that integrates state-of-the-art image segmentation with foundation model-based feature engineering and attention-based multiple instance learning (ABMIL) to predict EGFR mutation status from the histopathology WSI (Ilse et al., 2018; Vaidya et al., 2025; Xu et al., 2024; Zhang et al., 2025). To the best of our knowledge, our study is the first of its kind where foundation models-based feature embedding and transfer learning were applied on an Indian cohort. Lastly, our pipeline has achieved   a higher AUC (0.933) as compared to previous study conducted on the Indian population by Gupta et al., 2022 (AUC = 0.865) (Gupta et al., 2023) and the latest study by Campanella et al., 2025 AUC of 0.847 (Campanella et al., 2025). The final model was also tested on an independent external test set from The Cancer Genome Atlas (TCGA) database, demonstrating robust generalizability, irrespective of the training domain and imaging modality.

**Materials and Methods**

*1. Slides Selection and Screening Criteria*

This investigation constituted a retrospective cohort study, duly conducted following the institutional ethics guidelines. A manual computer-based data search was initially performed on the in-house databases of the; namely ePath and LIS, to extract a comprehensive list of all cases previously diagnosed with adenocarcinoma of the lung or Non-Small Cell Lung Cancer - Not Otherwise Specified (NSCLC-NOS). These were guided by molecular testing results, to establish a cohort comprising at least 110 cases of EGFR-mutated NSCLC, 60 cases of ALK-rearranged NSCLC, and 20 cases of ROS1-rearranged NSCLC. Additionally, a cohort of cases negative for EGFR mutations, ALK alterations, and ROS1 alterations (hereinafter referred to as triple-negative cases) was incorporated to ensure a balanced dataset that accurately represents the spectrum of major molecular subtypes of NSCLC (Table 1). The Hematoxylin and Eosin (H&E) stained slides of all shortlisted cases. These slides encompassed both biopsy and resection specimens from the selected patients. All slides underwent review by pathologists (DJ, SS) to verify the diagnosis of NSCLC and to ensure the adequacy of tumour content. Slides comprising a minimum 20% tumour cellularity, optimal H&E staining characteristics.

*Table 1. Dataset summary showing WSI used for model development and evaluation.*

| Class | EGFR +ve | EGFR -ve | | | Total |
|---|---|---|---|---|---|
| Variants | EGFR mutation | ALK mutation | ROS mutation | Triple -ve (ALK, ROS & EGFR wt) | |
| Train/Val | 94 | 54 | 16 | 6 | 170 |
| Test | 16 | 6 | 4 | 4 | 30 |
| Total | 110 | 60 | 20 | 10 | 200 |

## 2. Whole Slide Imaging

All 200 H&E-stained slides were scanned utilizing the Nanozoomer S60 digital slide scanner (Hamamatsu Photonics, Japan). The operational conditions were standardized with a scanning resolution of 40× magnification (0.23 μm/pixel) and a focus mode employing 3-point autofocus for each slide. Z-stacking was performed in three layers with two-micron intervals, while real-time image quality assessment facilitated quality control. The scanning procedure adhered to rigorous protocols to ensure consistent image quality. Each slide was meticulously cleaned with lint-free wipes to eliminate dust particles before being loaded into the scanner's autoloader in batches of 60. A pre-scan quality check was executed to detect potential scanning issues, which were simultaneously rectified if necessary. Full-slide scanning commenced with automated tissue detection, supplemented by a manual post-scan quality assessment to identify any focusing errors or artefacts. Any encountered errors were rectified at the source. The resultant digital whole slide images (WSIs) were saved in the proprietary 'ndpi' (NanoZoomer Digital Pathology Image) format, preserving the hierarchical multi-resolution structure of whole slide images. Each digital slide contained metadata detailing scan parameters, timestamps, and a unique identifier associated with case information, ensuring patient confidentiality was upheld.

### 3. WSI Preprocessing and Feature Embedding

The dataset comprising 200 whole-slide images (WSI), were divided into training (85%) and test (15%) sets using stratified sampling approach to preserve the overall class distribution across both datasets (Table 1). Trident pipeline was used to perform background noise (non-informative regions e.g. glass slide margin, white space, air bubble, etc.) removal and segmenting each WSI into non-overlapping patches of $256 \times 256$ pixel (Vaidya et al., 2025; Zhang et al., 2025). Subsequently, feature extraction was performed on generated patches (approx. 2,000 patches each WSI) using GigaPath vision transformer (Xu et al., 2024). The embedding vector generated from the Trident integrated Prov-GigaPath encoder was stored in .h5 file format and further used for downstream deep-learning (DL) analysis.

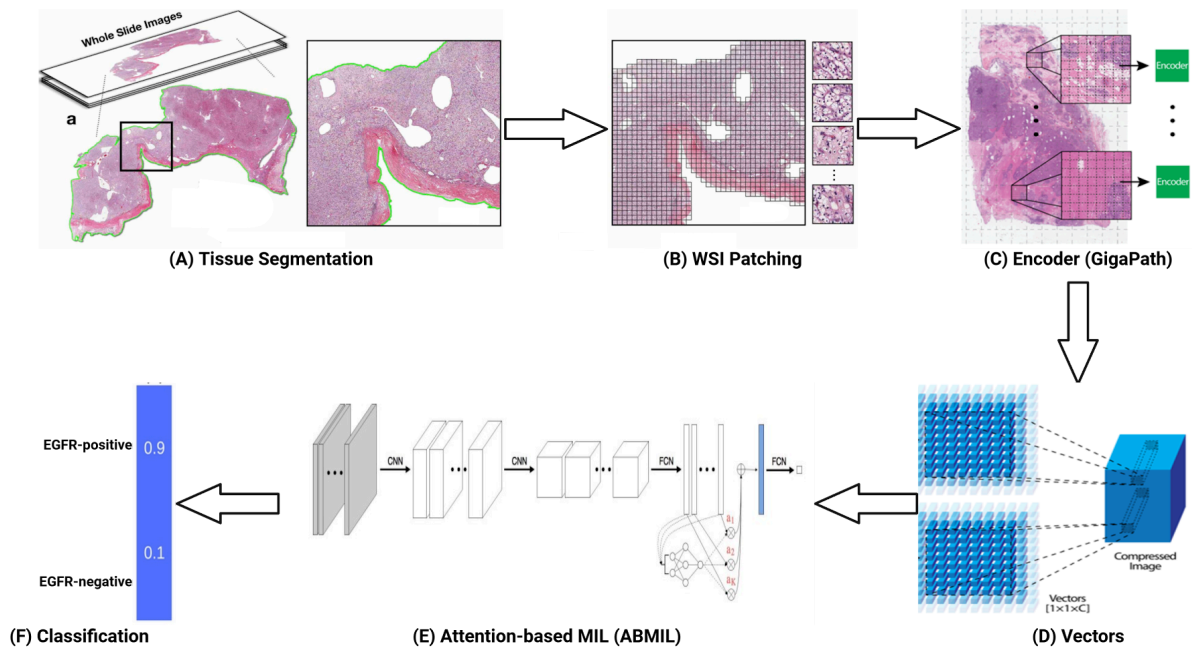

*Figure 1. Overview of the proposed framework for predicting EGFR mutation status from WSI.*

### 4. Predicting EGFR Mutations status

The pretrained ABMIL model was fine-tuned using 5-fold cross-validation (CV) on the training set (Indian dataset). In each fold, the data were randomly partitioned into five subsets, with four used for training and the remaining one for validation. This was done to ensure robust and unbiased model training and evaluation across the entire dataset. We applied an early stopping strategy to prevent overfitting and make the training loop more efficient. If the model's validation loss didn't improve for 8 consecutive epochs, training was stopped (Prechelt, 2012). The model with the highest cross-validation accuracy was selected and evaluated on the held-out test set to access its true performance. To further access its generalizability, the final model was evaluated on an external test set from TCGA with 86 WSI images (i.e. 43 each for EGFR+ and EGFR-class). The model's performance was assessed using a range of evaluation metrics, including accuracy, precision, recall, F1-score, area under the receiver operating characteristic curve (AUC), Matthew's correlation coefficient (MCC), area under the receiver operating characteristic (AUC-ROC) curve and confusion-matrix. For all the analysis (training and evaluation), WSI with single EGFR mutation status was selected corresponding to EGFR+ class, and images with multi-mutations were excluded.

All the analysis was performed on a high-performance Linux workstation (Ubuntu 18.04.6 LTS) with 512 GB RAM, 64-core CPU, and an NVIDIA RTX A5000 GPU (24 GB VRAM) with CUDA 12.4 support. The entire workflow was implemented in Python with the Pytorch framework.

**Results**

The proposed framework i.e. optimized preprocessing pipeline with fine-tuned classification model (on Indian dataset) was evaluated on two independent datasets: internal (Indian dataset) and external (TCGA dataset) test set. For both the datasets, WSI were subjected to the optimized preprocessing pipeline, as depicted in Figure 1. This pipeline included tissue segmentation, tile extraction and feature encoding. Finally, EGFR mutation status was predicted using the ABMIL model, classifying WSI into one of two categories: EGFR-positive (EGFR+), indicating the presence of an EGFR mutation, or EGFR-negative (EGFR−), indicating wild-type EGFR status. The optimized workflow enables consistent and reproducible analysis of WSI across different datasets and ensures reliable prediction of EGFR mutation status. On the internal test set consisting of WSI from the Indian population, the model achieved an AUC of 0.933 ± 0.010 suggesting excellent discriminative power between EGFR+ and EGFR- classes. The F1-score was 0.875 ± 0.019 indicating a strong balance between precision and recall. Additionally, the MCC, which accounts for both false positives and false negatives and is especially informative for binary classification tasks, was 0.732 ± 0.014 (as shown in Table 2).

*Table 2. Model performance for predicting EGFR mutation on different WSI datasets (Indian and TCGA).*

| Dataset | F1-Score | AUC | MCC |
|---|---|---|---|
| **Internal Test (Indian)** | 0.875 ± 0.019 | 0.933 ± 0.010 | 0.732 ± 0.014 |
| **External Test (TCGA)** | 0.963 ± 0.016 | 0.965 ± 0.015 | 0.932 ± 0.014 |

Figure 2 depicts the AUC-ROC curve and confusion-matrix plots, providing a visual assessment of the model's performance on the internal test set. The model performance on the internal test set was comparatively low as compared to cross-validation i.e. AUC of 0.969 ± 0.015, F1-score of 0.932 ± 0.027 and MCC 0.845 ± 0.062. The drop in performance is generally seen when transitioning from validation to test data and underscores the model's robustness on unseen samples.

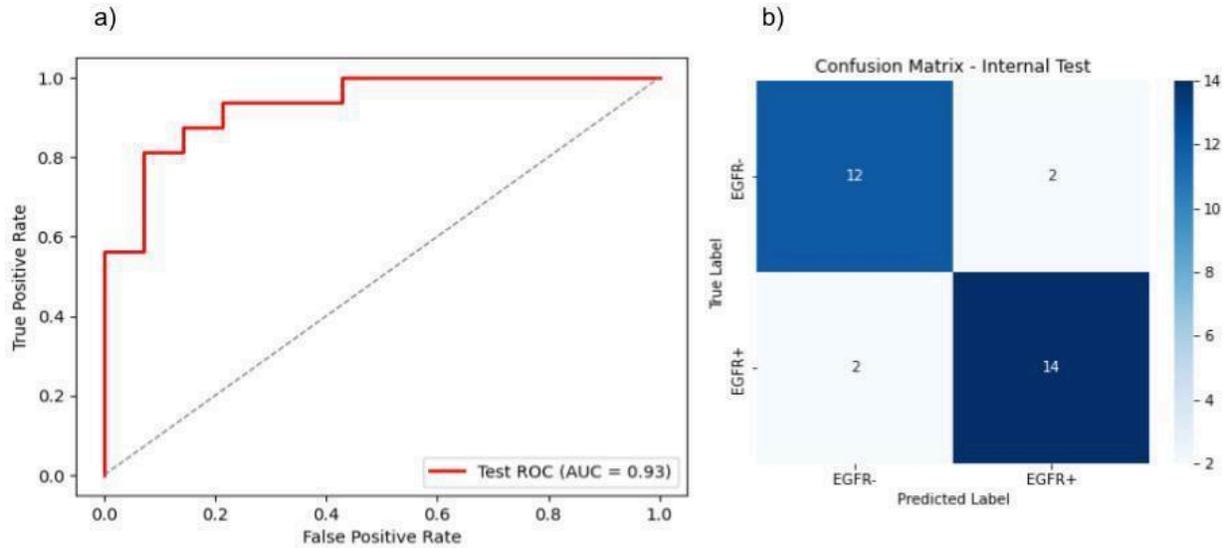

*Figure 2. AUC-ROC curve (a) and confusion-matrix (b) showing performance on the internal test (held-out) set from the Indian population.*

Finally, to assess the real-world applicability and generalization capability of the optimized framework, we evaluated it on a completely independent external cohort from TCGA, which has different domain distribution in terms of patient demographics, slide preparation, and scanner characteristics. Remarkably, the model not only retained its performance but exceeded expectations, achieving an AUC of 0.965 ± 0.015, F1-score of 0.963 ± 0.016 and MCC of 0.932 ± 0.014 (as shown in Table 2). The superior performance on external test sets can be due to the fact that encoders such as Prov-GigaPath are benchmarked and optimized to perform well on public datasets like TCGA.

We also implemented the model (EAGLE) developed by Campanella et al., 2025 using custom python scripts. The model was evaluated on the 200 Indian LUAD WSI dataset (as described in Table 1). On the Indian dataset, EAGLE achieved an AUC of 0.62 (Figure 3), which is significantly lower than the AUC of 0.847 reported in the original study. This highlights that the model is underperforming on the Indian dataset, potentially due to limited representation of Indian-specific data in the training and validation cohort

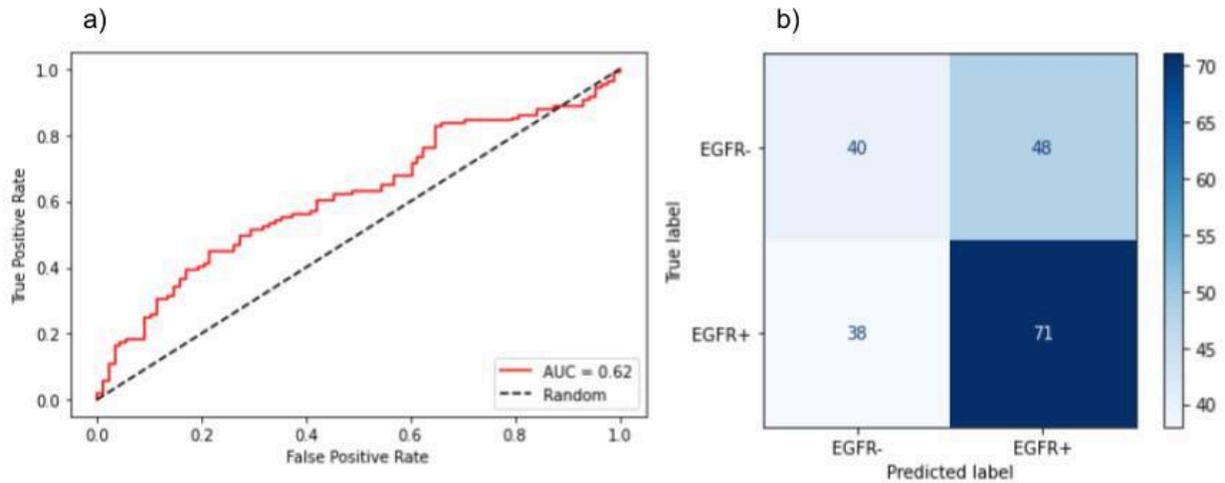

*Figure 3. AUC-ROC curve (a) and confusion-matrix (b) showing the performance of EAGLE on the Indian dataset.*

## Discussion

The present study proposes an efficient and optimized DL framework to predict EGFR mutation status directly from hematoxylin and eosin (H&E)-stained WSI. The aim of the study was to develop a robust preprocessing pipeline integrated with a classification model capable of predicting actionable mutation from routinely available pathology slides without requiring additional molecular testing (NGS) as input. The preprocessing pipeline implemented in this study is state of the art and utilizes Trident toolkit and Prov-GigaPath a ViT-based pathology foundation model. The EGFR prediction ABMIL model was fine-tuned using WSI from the Indian population. We tested this framework across two independent datasets: an internal (Indian) held-out test set, and an external (TCGA) test set. Across all test sets, the model consistently shows high performance, underscoring its robustness, reliability, and capacity to generalize to unseen data.

This is the first study to apply the ViT-based foundation model and an attention-based learning strategy to predict EGFR mutations in a cohort from India. Most prior studies in this field have primarily focused on Western populations, and there remains a substantial gap when it comes to AI model development and validation on data from Indian settings. Our model helps address this imbalance and contributes valuable evidence to the growing field of AI enabled digital-pathology. Focusing on its performance, the results are compelling. The model achieved an AUC of 0.933 (± 0.010) on the internal (Indian) test set, and an impressive 0.965 (± 0.015) on the external (TCGA) test set. These consistent results across different datasets strongly suggest that the model has learned meaningful patterns and isn't overfitting the training data.

The proposed preprocessing pipeline combined with ABMIL architecture plays a key role in achieving this performance. Unlike traditional AI enabled digital-pathology pipelines the focus of the current study was on optimum feature embedding using ViT foundation model. The embedded feature vector was then imputed to the EGFR prediction model. This is particularly important in histopathology, where relevant diagnostic features may be spatially dispersed. By leveraging trainable attention mechanisms, the ABMIL model learns to focus on regions that are most informative for classification, effectively mimicking the way a pathologist might concentrate on specific areas of a tissue slide.

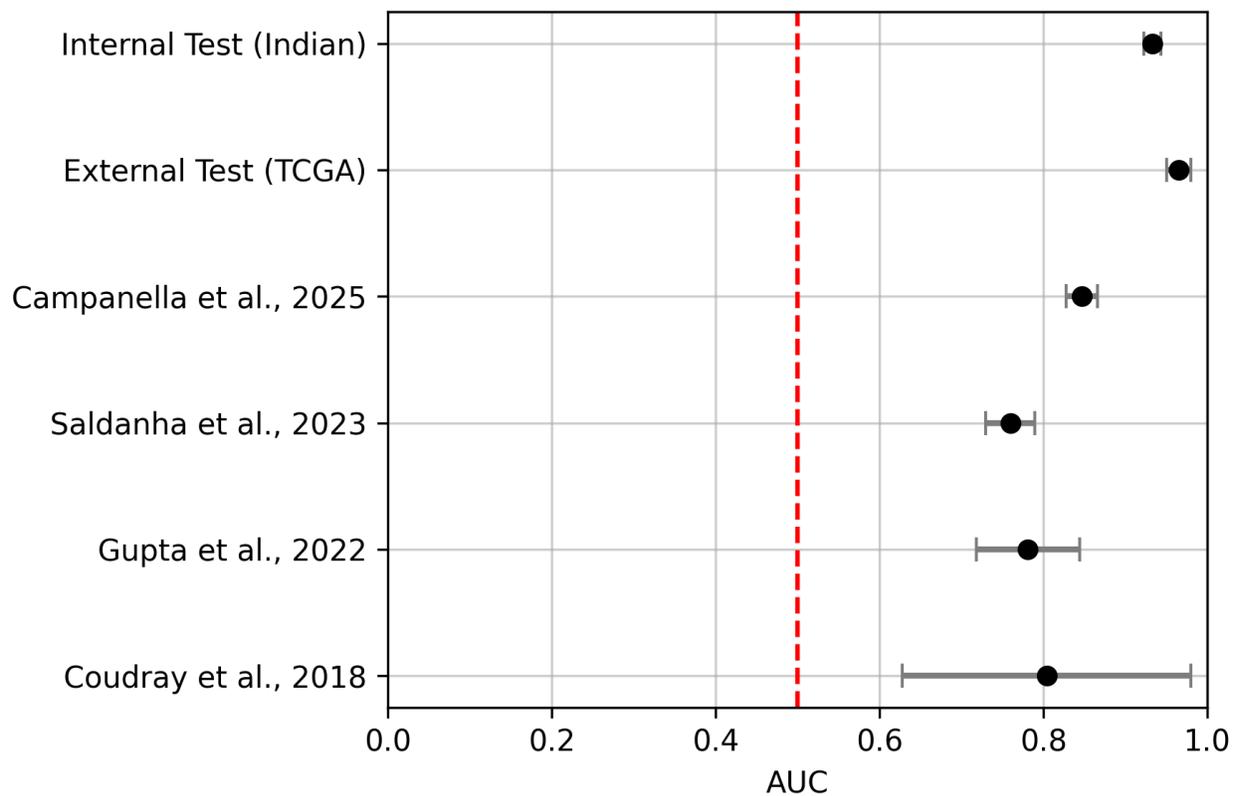

*Figure 4. Forest plot comparing AUC values for EGFR mutation prediction between this and previous studies.*

We have also compared the performance of our framework with findings from previous studies in the same domain. Figure 4 presents a forest plot of AUC values reported in past literature alongside our model's performance (internal and external test). Our method outperforms earlier models like that of Coudray et al., 2018, which reported an AUC of 0.804, and Gupta et al., 2022, which achieved 0.781 on an Indian dataset. More recent efforts, such as those by Saldanha et al., 2023, and Campanella et al., 2025, also fall short of the performance demonstrated in our study (Campanella et al., 2025; Coudray et al., 2018; Gupta et al., 2023; Saldanha et al., 2023). This comparative edge indicates that ViT-enabled feature embedding and attention-based model architectures may offer significant advantages over traditional or even more recent approaches when applied to histological data. Furthermore, we attempted to benchmark the model proposed by Campanella et al., 2025 on the Indian dataset; however, we were unable to successfully set it up using the available code on GitHub and HuggingFace. This issue has also been reported on the official GitHub repository. Consequently, we implemented the model using custom python scripts. On the Indian dataset, the model achieved an AUC of 0.62, which is significantly lower than the reported AUC of 0.847. These findings further underscores the need to develop AI models that are specifically trained or adapted for the Indian population.

Introducing AI enabled digital-pathology for clinical decision making can raise critical considerations around model explainability, ethics and utility. While our framework shows great potential, we don't view it as a replacement for trained pathologists. Instead, it should be looked at as a supportive technology for pathologists and oncologists to prioritize cases, especially in resource constrained clinical settings. In developing countries like India with high cancer burden, extremely low pathologist per million population, long turn-around time for genetic testing that may be even unaffordable leads to significant loss to follow-up of patients. Our framework can be easily deployed in such a scenario to identify and stratify patients with high risk of actionable mutations (e.g. EGFR mutation) from histopathological WSI. Such triaging can ensure better accessibility of resources for patients, leading to timely treatment interventions. One of the key strengths of our framework is that it can be trained on relatively small datasets, allowing for easy adoption in smaller clinical settings where computational resources are limited and large-scale model training isn't practical.

The proposed framework has been validated on a smaller dataset collected from a single center and doesn't provide support for model explainability. Looking ahead, we plan to build on this work by implementing a multi-centric approach for data collection and incorporating other actionable mutations. We are also exploring techniques to further improve domain generalization and implement model interpretability tools for real-world deployment. Another important next step would involve validating the model through prospective clinical studies, to understand how it performs in day-to-day diagnostic workflows.

**Conclusion**

In this study, we present a robust and optimized DL framework for predicting EGFR mutation status directly from H&E-stained WSI using the ViT and ABMIL model. The model has achieved consistently high performance across internal and external datasets, outperforming several established benchmarks for EGFR mutation prediction. Our approach holds particular promise for resource constrained settings with a huge patient burden and low pathologist to patient ratio as in India. The proposed framework requires only routinely available pathology slides and can be trained and validated on relatively small datasets (as small as 200 slides), which makes it highly suitable for deployment in smaller clinical centers or rural hospitals settings where molecular diagnostics may not be available, leading to loss to follow-up of patients. By prioritizing high risk patients with EGFR mutation, this model can improve diagnostic efficiency and timely treatment interventions. Looking forward, we aim to expand this framework to support multiclass mutation prediction, validate its performance in prospective clinical trials, and explore further improvements in domain adaptation and model explainability. Overall, our findings highlight the potential of attention-based DL models integrated with advanced ViT-based pathology foundation models to bridge the gap between advanced diagnostics and real-world clinical accessibility, supporting the broader goal of personalized medicine in lung cancer care.

## Funding

The study was partially supported by the ICMR grant IIRP-2023-2341, DBT Ramalingaswami fellowship, and IITD intramural MFIRP grant MI02999 to IG and DHR-ICMR (DIAMOnDS) project code I-1180 to DJ. This work is supported by the Central Research Facility (CRF) and High-Performance Computing (HPC) facility at the Indian Institute of Technology (IIT) Delhi.

## Author Contributions

IG, DJ, and PSM conceptualized the study. IG and DJ provided funding. SSG and R performed all the analysis and wrote the final manuscript with the help of IG and DJ. DJ, SD, SS, NMI, and VJ were responsible for data collection and annotation. AG provided critical input during data analysis. All authors contributed and approved the final manuscript.

## Code and Data Availability

The codes and model weights can be provided on reasonable request to the corresponding authors. The internal dataset (Indian dataset i.e. 200 WSI and corresponding clinical metadata) cannot be provided as per the institution policy. External TCGA dataset can be downloaded from https://portal.gdc.cancer.gov/projects/TCGA-LUAD.